\documentclass[12pt]{article}

\usepackage[hyperfootnotes=false]{hyperref}
\usepackage[hyperfootnotes=false]{hyperref}

\usepackage{mathtools}
     \usepackage{amsmath}
      \usepackage{amssymb}
  \usepackage{graphicx}
  \usepackage{xcolor}
  \usepackage{dsfont}
  \usepackage{physics}
  \newlength{\abstractwidth}
 \usepackage{mciteplus} 
\usepackage{subcaption}
 \usepackage{bbold}

  \newcommand{\be}{\begin{equation}}
  \newcommand{\ee}{\end{equation}}
\newcommand{\bea}{\begin{aligned}}
\newcommand{\eea}{\end{aligned}}
  \newcommand{\p}{\partial}

\newcommand{\bz}{\bar{z}}

\newcommand{\vepsilon}{\varepsilon}
\newcommand{\calM}{\mathcal M}
\newcommand{\calI}{\mathcal I}
\newcommand{\calZ}{\mathcal Z}

\newcommand{\calK}{\mathcal K}

\newcommand{\wtilde}{\widetilde}

\newcommand{\mfT}{\mathfrak T}

\DeclareMathOperator{\Sch}{Sch}
\DeclareMathOperator{\const}{const}

\DeclareMathOperator{\AAdS}{AAdS}

\DeclareMathOperator{\TFD}{TFD}
\DeclareMathOperator{\onshell}{on-shell}

\DeclareMathOperator{\EE}{EE}

\def\32{{3 \over 2 } }

  \def\ba{\begin{eqnarray}}
  \def\ea{\end{eqnarray}}

 \def\simleq{\; \raise0.3ex\hbox{$<$\kern-0.75em
      \raise-1.1ex\hbox{$\sim$}}\; }
 \def\simgeq{\; \raise0.3ex\hbox{$>$\kern-0.75em
      \raise-1.1ex\hbox{$\sim$}}\; }
\title{On thermofield-double's effective degrees of freedom in three dimensions}

\author{Pouria Dadras\footnote{pdadras@caltech.edu, pouria.dadras@gmail.com}\\
\normalsize\it California Institute of Technology, Pasadena, CA 91125, U.S.A. \\\\ \normalsize\it School of physics, Institute for research in fundamental sciences, \\ \normalsize\it P.O.Box 19395-5531, Tehran, Iran
\vspace{0.5cm}}

\date{}

\begin{document}

\setcounter{tocdepth}{2}

\maketitle
\begin{abstract}
 We will give two representations for thermofield-double (TFD). The first representation is the well-known Kerr-BTZ black hole geometry, which is the solution of Einstein's equation. The second representation is a disjoint union of two circles, which is a solution to two copies of the Schwarzian theory. This representation, in particular, admits the reparametrization modes, which are absent in any two dimensional CFT on a torus, and are essential for (dis-)entangling the TFD. Since the two representations describe the same state, they must be equivalent.   

\end{abstract}
\tableofcontents
\newpage
\section{Introduction and Outline of the paper \label{introduction}}
The mysterious nature of an eternal black hole has offered many  motivations for physicists to investigate its geometry in the past decades. While the metric can be considered as one of the simplest solutions to the vacuum Einstein equation, it has non-trivial features. The existence of the event horizon, a codimension one null hypersurface where according to an outside observer, nothing can enter or escape from it, the existence of a second side in the maximal extension of the geometry, the fact that its parameters can be identified with thermodynamic quantities that satisfy the laws of thermodynamics \cite{Hawking73}, and last but not least, associating an entangled state, namely the thermofield-double state \cite{Israel76} to the geometry are among such features. These surprising properties have had a significant role in the development of  proposals, such as the  Holographic principle \cite{Hooft93,Susskind95}, entanglement structure of spacetime \cite{Raamsdonk09}, and ER=EPR \cite{Maldacena13}. \\
The AdS/CFT duality \cite{Maldacena99,Witten98,Gubser98} is one of the  systematic frameworks for investigating black holes. One of the consequences of the duality is that for certain ``holographic theories,'' correlation functions of the operators inserted on the boundary can also be computed from bulk using the AdS/CFT dictionary. While finding the exact holographic theory can be challenging, investigating the bulk puts some strong constraints on it. One of the constraints is the MSS bound \cite{Maldacena15} based on the earlier work \cite{Shenker13,Shenker14}, which implies that the holographic theory should satisfy the saturation of the chaos bound as the theory at finite temperature should be dual to a bifurcate horizon in the bulk. \\
The Sachdev-Ye-Kitaev model \cite{Sachdev93,Kitaev17,Maldacena16} is an example of a theory that meets this criterion. In its simplest form, the theory consists of a large number of Majorana fermions coupled by a random coupling. At low energy limit, the theory has an emergent reparametrization symmetry whose dynamics are effectively described by the Schwarzian action. Such modes are responsible for the saturation of the chaos bound. Moreover, the action has a geometric representation; it can be derived from the GHY boundary action in the Jackiw-Teitelboim (JT) gravity \cite{Jackiw84,Teitelboim83} in the limit where the boundary of AdS2 space has small fluctuations \cite{Maldacena16a}. \\ In holography, $\ket{\text{TFD}}$, represented by 
\be
\ket{\TFD}= \frac{1}{\calZ^\frac{1}{2}} ~ \sum_n ~ e^{-\frac{1}{2}\beta ( E_n- \Omega_H  J_n)}\ket{n_R} \otimes \ket{n_L},
 \ee
 is conjectured to be dual to the AdS-Kerr black hole \cite{Maldacena03}. Here, $E_n,J_n$ are the eigenvalues of the Hamiltonian and the angular momentum operators, the generators of time and angular directions on the boundary of AdS, whose expectation values are identified with the ADM energy and angular momentum of the black hole in the bulk. In holographic theories, there are also other emergent directions in the bulk. We can then compute the associated partition function using the Hawking-Gibbons' prescription \cite{Gibbons76,Hawking83}. For the Kerr-BTZ black hole \cite{Banados92,Banados93}, See \cite{Carlip98} for a review, \be
\ln \calZ = \tr (e^{-\beta (H-\Omega_H J)}) = \frac{\text{entropy}}{2}, 
 \ee
 where in above by ''entropy'' we mean the area of the black hole horizon divided by $4G$. \\\\ There have been efforts to find a comprehensible quantum theory of gravity. Even in three dimensions, where the gravitational degrees of freedom are absent, computing the partition function with a given boundary condition seems to be challenging \cite{Witten88,Witten07,Witten10}.\\\\
 In this paper, we are not attempting to compute the partition function of quantum gravity in three dimensions. What we will propose is that, in addition to the Kerr-BTZ black hole, it is possible to associate a second geometric representation to TFD, namely, disjoint union of two circles whose fluctuations are described by two copies of the Schwarzian action. The partition function is now, 
 \be
\ln \calZ = -\frac{1}{8G}\Big(\Sch_+ + \Sch_- \Big) = \frac{\text{entropy}}{2}, 
 \ee
 where $\Sch_{\pm}$ are the value of the Schwarzian action over the classical solution.
 Microscopically, these circles may be identified with the ''Euclidean evolution'' of two SYK models. \\\\
 From field theory point of view, as long as one is interested in gravitational aspects of the Kerr-BTZ black hole, computations involving two copies of Schwarzian (the second representation of TFD) is preferred over doing quantum field theory on a two dimensional torus. In fact, on two circles, the reparametrization modes live, which are responsible for out of time ordered correlators (OTOC)s with maximum Lyapunov exponents. Moreover, as was shown in\cite{Dadras22}, by coupling left and right side of TFD \footnote{This setup was originally used\cite{Gao16}to make the Einstein-Rosen bridge in the bulk traversable. See \cite{Maldacena18, Chen19} for a similar setup.} and evolving the system, the entanglement entropy will change. For holographic theories, the system equilibrates fast enough but with a different temperature, which can be read from the two-point function. The modes responsible for the temperature change must be generic, i.e. they should be independent of the UV degrees of freedom. In fact, such modes are the reparametrization modes. Note that these modes are absent in a torus topology due to Liouville's theorem, which mentions that the only bounded and holomorphic function is constant. \\\\
 
 In section \ref{section2}, we will first give a brief review of the Kerr-BTZ black hole. We will then propose the second representation of $\TFD$. Section \ref{section3} will be a summary of results in \cite{Maldacena16}, \cite{Kitaev17} and \cite{Dadras22}, which show why two copies of Schwarzian action may be a right substitution for a two dimensional CFT. In section \ref{section4}, we will propose an intuitive way to relate these two representations. There is also an appendix which studies the Hawking-Gibbons-York boundary action close to the boundary of AdS$_3$.       
 \section{The two representations of the thermofield-double \label{section2}}
 For a many body quantum mechanical system, thermofield-double has the following form
 \be \label{tfd1}
\ket{\TFD} = \frac{1}{\sqrt{\calZ}} ~ e^{-\frac{\beta}{2}(\hat{H}_R - \Omega \hat J_R)} \sum_n ~~ \ket{n_L} \otimes \ket{n_R},
\ee
where $\ket{n_R}$ are the mutual eigenstates of $H_R$ and $J_R$, and
\be
[\hat H_R,\hat J_R] = 0,~~~~~~\hat H_R \ket{n_R} = E_n \ket{n_R},~~~\hat J_R \ket{n_R} = J_n \ket{n_R}. \ee We further assume that $~~0\le|J_n| \le E_n $. \subsection{The first representation}
  
 The Kerr-BTZ geometry in the standard coordinate system is given by the following metric:
\be \label{BTZ}
\begin{aligned} &
ds^2 = -\frac{(r^2-r_+^2)(r^2-r_-^2)}{r^2} \, dt^2 + \frac{r^2}{(r^2-r_+^2)(r^2-r_-^2)} \, dr^2 +r^2 \Big( \frac{r_- r_+}{r^2} dt - d\phi \Big)^2,
	\end{aligned}
\ee
which is identified by two parameters $r_+$ and $r_-$, the inner and outer horizon of the black hole,  or equivalently, the ADM mass and angular momentum, which are related by 
\be
 M_{ADM} = \frac{r_+^2+r_-^2}{8G},~~~~~~J_{ADM} = \frac{r_+r_-}{4G}.
\ee
The inverse temperature and the horizon angular velocity are given by:
\be
\beta_H = \frac{2\pi r_+}{r_+^2-r_-^2},~~~~~~ \Omega_H = \frac{r_-}{r_+}.
\ee
Since the Weyl tensor vanishes identically in three dimensions, it is possible to derive the metric \ref{BTZ} by a pure coordinate transformation of the Poincare half-plane metric:
\be
ds^2 = \frac{dR^2 - dU_-dU_+}{R^2}
\ee
\be
\begin{aligned} &
R = \Big(\frac{r_+^2-r_-^2}{r^2-r_-^2} \Big)^{\frac{1}{2}} ~ e^{-r_- t +r_+\phi},~~~U_{-} = \Big(\frac{r^2-r_+^2}{r^2-r_-^2} \Big)^{\frac{1}{2}} ~ e^{\frac{2\pi}{\beta_{-}}(t +\phi)},~~~U_+ = -\Big(\frac{r^2-r_+^2}{r^2-r_-^2} \Big)^{\frac{1}{2}} ~ e^{-\frac{2\pi}{\beta_{-}}(t -\phi)} \\&
~~~~~~~~~~~~~~~~~~~~~~~~~~~~~~~~~~~~~~~~~~~~~~~~~~~~~~~~~\beta_{\pm} = \frac{2\pi}{r_+ \pm r_-},	\end{aligned}
\ee
\begin{figure} [t]
\centering
\includegraphics[scale=.8]{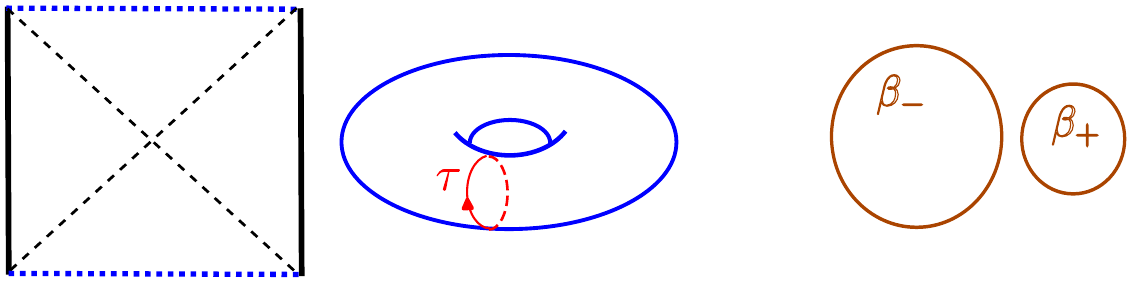}
\caption{The two representation of TFD. The left hand side depicts the Lorentzian and Euclidean Kerr-BTZ, while the right side is the disjoint union of two circles.}
\label{fig:tworep}
\end{figure}
with an additional constraint $\phi \sim \phi+2\pi$. As a part of identifying \ref{tfd1} and \ref{BTZ}, one may set $\hat H$ and $\hat J$ as generators of the time and angular translations, which are symmetries of the metric, with the expectation value set to  
\be
\langle \hat H_R \rangle = M_{ADM},~~~~\langle \hat J_R \rangle = J_{ADM},~~~~\Omega = \Omega_H,~~~~\beta = \beta_H\ee
 To compute the partition function, following Hawking and Gibbons, we need to compute the Einstein-Hilbert action with Hawking-Gibbons-York boundary term over the Euclidean BTZ, which corresponds to $t=-i\tau$ and $r_- = i|r_-|$ with the following identification:
\be
(\,\tau \,, \,\phi \,) \sim (\, \tau+\beta_H \,, \, \phi +i \beta_H \, \Omega_H \,).
\ee
 It is more convenient to define $z = \tau+i\phi$ and $\bz = \tau - i\phi$. Then we have 
\be \label{torus}
( \,z \,, \,\bz \,) \sim (\, z+\beta_+ \,, \, \bz +  \beta_-),~~~~~~~ \beta_+ = \bar \beta_-. \ee
The geometry in Euclidean signature describes a torus with the modular parameter $\tau = \frac{\beta_+}{2\pi}$. We then identify the partition function and Euclidean action through 
\be \label{onshell}
-\ln \tr (e^{-\beta (\hat H-\Omega \hat J)}) \approx S_{\onshell} = \beta M -S - \beta \Omega J = -\frac{\pi r_+}{4G}.
\ee
While clear, It is important to have in mind that the Einstein-Hilbert term and especially, the black hole horizon always has a nonzero contribution to the on-shell action. 
\subsection{The second representation}
In this representation we first define $\hat H_{\pm} = \frac{1}{2}\Big(\hat H\pm \hat J\Big)$. Note that $\hat H_\pm$ are generators along the causal directions $t\mp \phi$. Rewriting the thermofield-double in terms of the new variables, we will obtain
\be
\begin{aligned} &
\ket{\TFD} = \frac{1}{\sqrt{\calZ}}~ e^{-\frac{\beta_+ \hat H_++\beta_- \hat H_-}{2}} \sum_{n_+, n_-} \ket{E_{n_+},E_{n_-}}_L \otimes \ket{E_{n_+},E_{n_-}}_R = \ket{TFD}_+ \otimes \ket{TFD}_- \\&
\ket{\TFD}_{\pm} = \frac{1}{\sqrt{\calZ_{\pm}}} ~e^{-\frac{\beta_{\pm}}{2} \hat H_{\pm}}\sum_{n_{\pm}} \ket{E_{n_\pm}}_L \otimes \ket{E_{n_\pm}}_R.
\end{aligned}\ee
Notice that the partition function factorizes into the partition function of two canonical ensembles with inverse temperatures $\beta_{\pm}$,
\be
\ln \calZ = \ln \calZ_+ (\beta_+) + \ln \calZ_- (\beta_-),
\ee
which is similar to holomorphic factorization except that, here, $\beta_{\pm}$ are real. We also have
\be
\langle \hat H_{\pm} \rangle = \frac{1}{2} \Big(M_{ADM} \pm J_{ADM}\Big) = \frac{\pi^2}{4G} \frac{1}{\beta_\pm^2}
\ee
We can also read the entropy $S_{\pm}$ from the first law,
\be
dS_{\pm} = \beta_{\pm} dE_{\pm} \rightarrow S_{\pm} = \frac{\pi^2}{2G} \frac{1}{\beta_{\pm}}
\ee
(This entropy can also be derived from Cardy formula  in AdS$_3$/CFT$_2$\cite{Strominger98}.)
Consequently, the free energy of the two systems are
\be
F_{\pm} = -\frac{\pi^2}{4G} \frac{1}{\beta_{\pm}^2}. 
\ee
Note that the above thermodynamic quantities match the thermodynamic quantities associated with two copies of Schwarzian action that was studied previously in the context of the SYK model. In Euclidean signature, the Schwarzian actions are defined over the disjoint union of two circles with circumference $\beta_{\pm}$,
\be \label{twoSchwarzian}
\begin{aligned}
	I_E =- \frac{1}{8G} \int_0^{\beta_+} ~ du_E ~ \Sch (e^{i\varphi_+})- \frac{1}{8G} \int_0^{\beta_-} ~ dv_E ~ \Sch (e^{i\varphi_-}). 
\end{aligned}
\ee
While the new representation yields the same thermodynamic quantities as the previous one, now, the reparametrization modes survive. Such modes play an essential role in the next section.
\section{Evolution of the thermofield-double state by coupling its both sides \label{section3}}

In the previous section, we showed that in the second representation of $\TFD$, the effective action consists of two copies of the Schwarzian action. Inspired by the SYK model one can think of the action as the low energy limit of a large N microscopic theory (for example, Majorana fermions in the case of the SYK model). We call these microscopic fields $\phi^i_+$   and $\phi_-^i$ with scaling dimensions $\Delta_+$ and $\Delta_-$, with two point functions equal to
\be
\langle \phi^i_\pm(\theta^\pm_1) \phi^j_\pm(\theta^\pm_2) \rangle = \frac{(\varphi'(\theta^\pm_1))^{\Delta_\pm} (\varphi'(\theta^\pm_1))^{\Delta_\pm}}{\Big(\sin \frac{\varphi(\theta^\pm_1)-\varphi(\theta^\pm_2)}{2}\Big)^{2\Delta_{\pm}}} ~ \delta_{ij}. ~~~~~~0\le\theta <2\pi
\ee
where, $\varphi (\theta)$ is a reparametrization of $\theta$. Similar to the SYK model, we further assume that the contribution of the Schwarzian modes to the four point function dominates. Writing $G(\varphi(\theta_1), \varphi (\theta_2)) = G(\theta_1,\theta_2) \Big( 1 + \frac{\delta G(\theta_1,\theta_2)}{G(\theta_1,\theta_2)}\Big)$ in IR limit we will obtain \cite{Kitaev17,Maldacena03}
\be \label{t-ord} \begin{aligned} &
\bigg\langle \frac{\delta G(\theta_1^{\pm}, \theta_2^{\pm})}{G(\theta_1^{\pm},\theta_2^{\pm})} \frac{\delta G(\theta_3^{\pm}, \theta_4^{\pm})}{G(\theta_3^{\pm},\theta_4^{\pm})} \bigg \rangle = \frac{4 \, \Delta_{\pm1} \Delta_{\pm 2}}{S_{\pm}} \bigg( 1 - \frac{\frac{\theta_{1}^{\pm}-\theta_{2}^{\pm}}{2}}{\tan \frac{\theta_{1}^{\pm}-\theta_{2}^{\pm}}{2}} \bigg)\bigg( 1 - \frac{\frac{\theta_{3}^{\pm}-\theta_{4}^{\pm}}{2}}{\tan \frac{\theta_{3}^{\pm}-\theta_{4}^{\pm}}{2}} \bigg), \\& ~0\le\theta_1 < \theta_2< \theta_3 <\theta_4 <2\pi
\end{aligned}
\ee
for the case where fields with the same index are adjacent, i.e. $(iijj)$. In the case where the indices are alternating, $(ijij)$, the four point function will take the following form,
\be \label{ot-ord}
\begin{aligned} &
\bigg\langle \frac{\delta G(\theta_1^{\pm}, \theta_2^{\pm})}{G(\theta_1^{\pm},\theta_2^{\pm})} \frac{\delta G(\theta_3^{\pm}, \theta_4^{\pm})}{G(\theta_3^{\pm},\theta_4^{\pm})} \bigg \rangle = \frac{-\pi \sin \Delta \theta^{\pm}_c}{2\sin \frac{\theta^{\pm}_{12}}{2} \sin \frac{\theta^{\pm}_{34}}{2}} + \frac{-\pi (\pi - 2\Delta \theta^{\pm}_c)}{4\tan \frac{\theta^{\pm}_{12}}{2} \tan \frac{\theta^{\pm}_{34}}{2}} + \Big(1+\frac{\pi - \theta^{\pm}_{12}}{2\tan \frac{\theta_{12}^{\pm}}{2}}\Big)\Big(1+\frac{\pi - \theta^{\pm}_{34}}{2\tan \frac{\theta_{34}^{\pm}}{2}}\Big), \\&
0 \le \theta_1 < \theta_3 < \theta_2 < \theta_4 < 2\pi,~~~~~~\Delta \theta_c^\pm = \frac{\theta_1^\pm +\theta_2^\pm}{2} - \frac{\theta_3^\pm+\theta_4^\pm}{2}
\end{aligned} 
\ee
Now, we consider the following interaction unitary operator acting on thermofield-double,
\be
\ket{\wtilde{\TFD}} = U_I \ket{\TFD} = U^+_I \ket{\TFD}_+ \otimes ~U^-_{I} \ket{\TFD}_-,~~~~~~U^{\pm}_{I} = \mfT \exp \Big( -ig_{\pm} \int_0^{t_\pm} du ~ \phi_{\pm L} (-u) \phi_{\pm R} (u) \Big),
 \ee
 where $\phi_{\pm}$ are assumed to be Bosonic. As may be easily seen, the interaction Hamiltonian couples the two sides of the $\TFD_{\pm}$. However,  $\TFD_+$ and $\TFD_-$ remain unentangled. An important tool that we need to do the computation is the relation between $\phi_L$ and $\phi_R$ when acting on $\ket{\TFD}$. The $\ket{\TFD}_{\pm}$ is a cyclic and separating state with respect to the operator algebra generated by $\{\phi_{\pm R}\}$. Moreover, the operator algebra generated by $\phi_L$ and $\phi_{R}$ are each other's commutant. On acting on $\TFD_{\pm}$, they are related by Tomita's modular conjugation operator, which reads 
 \be \label{tt}
\phi_{\pm L}(u) \ket{\TFD_{\pm}} = \phi_{\pm R}(u+i \frac{\beta_{\pm}}{2}) \ket{\TFD_{\pm}}.
\ee
for the Bosonic operators and 
 \be \label{tt}
\phi_{\pm L}(u) \ket{\TFD_{\pm}} = i\phi_{\pm R}(u+i \frac{\beta_{\pm}}{2}) \ket{\TFD_{\pm}}.
\ee
for the Fermionic operators.  
  In this section we are interested in the dynamics of the entanglement entropy as well as thermodynamic quantities. Due to the fact that $\ket{\TFD_{\pm}}$ remain unentangled throughout the evolution, the problem breaks into the computation for two 0+1 dimensional systems. The computation was done in detail in \cite{Dadras22}, and we will only give a summary of the results. \\\\
$\bullet ~ $ \textbf{Evolution of entanglement entropy} \\\\
  The entanglement entropy can be computed using the replica method and relation \ref{tt}, where for two disjoint systems it is additive $S_{\EE} = S_{+\EE}+S_{-\EE}$. \be
  \begin{aligned} &
  \Delta S_{+\EE} = \frac{\pi g_+}{2J_+} \Big( \frac{\pi}{\beta_+ J_+}\Big)^{2\Delta_+-1} \Big( 1-\frac{1}{\Big(\cosh \frac{2\pi t_+}{\beta_+}\Big)^{2\Delta_+}} \Big) + \frac{\pi^2 g_+^2}{2S_+} \Big(\frac{\beta_+}{2\pi}\Big)^2 \Big( \frac{\pi}{\beta_+ J_+}\Big)^{4\Delta_+} \Bigg[\frac{4\Delta_+}{\cosh^{2\Delta_+} \frac{2\pi t_+}{\beta_+}} \times \\&\bigg( \frac{2\pi t_+}{\beta_+} \tanh \frac{2\pi t_+}{\beta_+}-1 +\frac{1}{\cosh^{2\Delta_+} \frac{2\pi t_+}{\beta_+}} +(2\Delta_+-1) \tanh \frac{2\pi t_+}{\beta_+} \int_0^{\frac{2\pi t_+}{\beta_+}} \frac{du}{\cosh^{2\Delta_+}u}\bigg) - \Big(1-\frac{1}{\cosh^{2\Delta_+} \frac{2\pi t_+}{\beta_+}} \Big)^2\Bigg] \\&
  ~~~~~~~+ O(g_+^3).
  \end{aligned} 
  \ee
  While in computing $S_{\EE}$ to leading order we only need to know the two point function, to second order the time ordered four point function must be known, and we used \ref{t-ord}. \\\\
  $\bullet ~ $ \textbf{Evolution of the Hamiltonian $H_{\pm}$} \\\\
  In computing the $\langle H_\pm \rangle$ to leading order in the coupling, we only need the two point function:
  \be
  \begin{aligned} &
\Delta H_{\pm} = \bra{\TFD} U_I^{-1}(t_{\pm}) \hat H_{\pm} U_I(t_{\pm}) \ket{\TFD} - \langle H_{\pm} \rangle = -ig_{\pm} ~ \int_0^{t_\pm} du ~ \langle [H_{\pm}, \phi_{\pm R}] ~ \phi_{\pm L} \rangle \\& = -g_{\pm} \int_0^{t_{\pm}}~ du ~ \langle \p_u \phi_{\pm R}(u) \phi_{\pm L} (-u) \rangle= \frac{g_{\pm}}{2} \Big(\frac{\pi}{\beta_\pm J_\pm}\Big)^{2\Delta_{\pm}} \Big(1 - \frac{1}{\cosh^{2\Delta_\pm} \frac{2\pi t_{\pm}}{\beta_{\pm}}} \Big) + O(g_\pm^2)
\end{aligned} 
  \ee
  It is clear that to leading order in $g_{\pm}$ we have the first law of thermodynamics,
  \be
\Delta H_{\pm} = T_{\pm} ~ \Delta S_{\pm\EE}.
  \ee
  $\bullet ~ $  \textbf{Evolution of two point function} \\\\
  We can also compute the evolution of two point function of two probing fields living on one side of  $\TFD$. In leading order in the coupling, we use the four point function \ref{t-ord} and \ref{ot-ord} and relation \ref{tt}. For $t_1,t_2 \simgeq \beta_{\pm}$ we have,
  \be
  \begin{aligned} &
\bra{\wtilde{\TFD}} O_{\pm R} (t_1) O_{\pm R}(t_2) \ket{\wtilde{\TFD}} \simeq G_{\beta_\pm}(t_2-t_1) \bigg(1+\frac{\beta_\pm g\Delta_\pm}{S_\pm (\frac{\beta_\pm J_\pm}{\pi})^{2\Delta_\pm}}\Big(1- \frac{\frac{\pi(t_2-t_1)}{\beta_{\pm}}}{\tanh \frac{\pi (t_2-t_1)}{\beta_{\pm}}} \Big)\bigg)~~~~~~ \\&~~~~~~~~~~~~ = G_{\wtilde \beta_\pm}(t_2-t_1),~~~~~\wtilde \beta_{\pm} = \beta_{\pm} \bigg(1- \frac{\pi g_{\pm}}{2J_{\pm} S_\pm} \Big( \frac{\pi}{\beta_\pm J_\pm}\Big)^{2\Delta_\pm-1} \bigg).
\end{aligned}
\ee
Notice that after the equilibrium the $SL(2,\mathbb{R}) \times SL(2,\mathbb{R})$ are revived, which confirms that $\wtilde{\TFD}_{\pm \beta}(t \gtrsim \beta_{\pm}) \rightarrow  \TFD_{\wtilde \beta_\pm }$ 
\section{A naive geometric relation between the two representations of $\TFD$ \label{section4}}
\begin{figure} [t]
\centering
\includegraphics[scale=1]{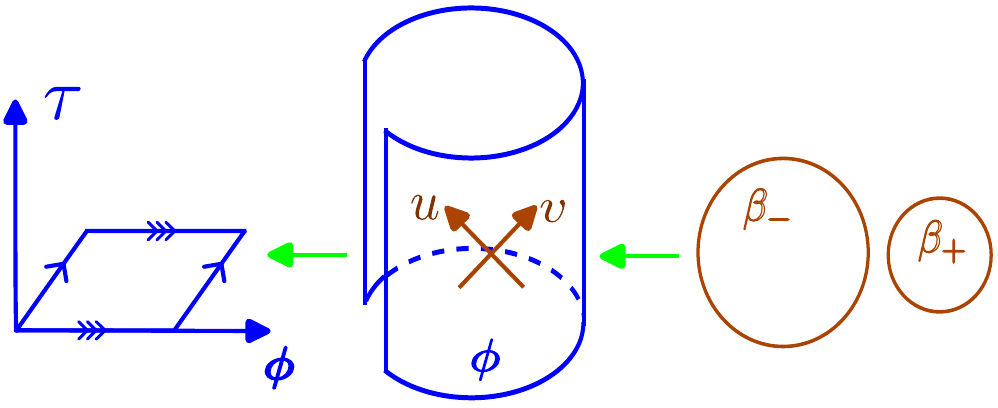}
\caption{}
\label{fig:cylinders}
\end{figure}

So far we have two representations for $\TFD$. In one representation, written in terms of the eigenstates of $\hat H$ and $\hat J$, a timelike and a spacelike generator, the associated geometry is the Euclidean BTZ black hole. On the other hand, the second representation of $\TFD$, written in terms of the eigenstates of $\hat H_\pm$ decouples, and is associated with two disjoint circles. Clearly, the BTZ representation has $U(1) \times U(1)$ symmetry, and also there is a $U(1)$ symmetry associated with each circle in the second representation. A possible way to relate these two pictures is to start from the classical solution $\varphi_+ = \frac{2\pi u}{\beta_+}$ and $\varphi_- = \frac{2\pi v}{\beta_-}$ of \ref{twoSchwarzian} in Lorentzian signature and attach $(u,v)$ directions to a two dimensional surface along the null directions, see figure \ref{fig:cylinders}. Note that the reprarametrization of the two lines are still possible.
We then take $\phi = \frac{v-u}{2}$ and identify $\phi$ with $\phi+2\pi$, which renders a cylinder. Since the $\phi$ direction is compactified, it is not possible to use the coordinate system $(u,v)$ globally for the cylinder. This implies that we cannot have both reparametrizations of $u$ and $v$ simultaneously. After the Wick rotation $t = -i\tau$, identifying the two ends of the cylinder by $\tau+i\phi \sim \tau + i\phi + \beta_+ $ (note that, here, $\beta_+$ is complex) we get a torus of modular parameter $\frac{\beta_+}{2\pi}$. In this new geometry, we will loose the remaining reparametrization modes. However, the torus is filled, and so the geometry is three dimensional. Following the above procedure, we can write a naive two dimensional area term by wedge producting $du$ and $dv$ with $d\phi$ and Wick rotating,
\be \label{naive}
I_E   \propto -\frac{1}{8 G} \int ~ d\tau \wedge d\phi ~\Big( \frac{a_+^2}{2} + \frac{a_-^2}{2} \Big),~~~~~~~~a_{\pm} = \frac{2\pi}{\beta_\pm}
\ee
where $\frac{a_+^2}{2}$ is the value of the Schwarzian over the classical solution.  We shall relate this to the Gibbons-Hawking-York boundary action. As is shown in appendix \ref{appendixA}, from the Hamiltonian and the momentum constraints, the trace of the second fundamental form associated with the boundary fluctuations for an asymptotic AdS$_3$ spacetime is
\be
\calK = -2 - \frac{1}{2} r^2 ~ ^2\wtilde R^{(0)} + O(r^3).  ~~~~~~\text{$r=0$ is the boundary}
  \ee
   If we write the black hole metric in the following form
  \be
ds^2 = \frac{-drdu-drdv-dudv}{r^2} + \frac{1}{4} a_+^2 du^2 + \frac{1}{4} a_-^2 dv^2 + \frac{1}{4} (a_+^2 + a_-^2) dudv,
  \ee
  the GHY action associated with the above metric is
  \be \label{GHY}
-\frac{1}{8\pi G} \int_{r=\vepsilon} ~ \Big( \calK+1 \Big) \sqrt{h_{E}} = \frac{1}{8\pi G} \int_{r=\vepsilon} d\tau \wedge d\phi ~ \Big( \frac{1}{2\vepsilon^2} -\frac{1}{8} (a_+^2+a_-^2)\Big),
  \ee
  and \ref{naive} and \ref{GHY} have the same form. Adding the Einstein-Hilbert action to \ref{GHY} yields the right value for the partition function. 
 \section{A short discussion and conclusion}
 In this article, we considered two representations for thermofield-double. The first representation, identified by eigenstates of the $\hat H$ and $\hat J$, corresponds to a filled torus with the time and the rotational symmetry. In Lorentzian signature, it corresponds to a Kerr-BTZ black hole. On the other hand, in the second representation, $\TFD$ is written as tensor product of two TFDs, identified by eigenstates of $\hat H \pm \hat J$. We then associated each TFD with a circle, and identified the corresponding partition function with the value of the Schwarzian action over these circles. Clearly, for this identification it is crucial to have $E_\pm \propto T^2_{\pm}$. \\\\
 The existence of reparametrization modes in the second representation of TFD is crucial for computations that are relevant to the gravitational picture. The full quantum free energy associated with such modes was computed in \cite{Stanford:2017thb}. The result is the free energy of the black hole plus a non-extensive logarithmic correction. Note that such modes are absent in any field theory on a two dimensional torus. This would also imply that excitations of the reparametrization modes in the second representation corresponds to sending matter into the bulk of the BTZ geometry. It would be interesting to understand the logarithmic correction from the gravitational representation. Note that the next correction to the Schwarzian action may be model dependent. For the case where the UV theory is the SYK model it is nonlocal \cite{Kitaev17}, and it is not clear whether they are  associated with geometry. 
 In an upcoming work we will more elaborate on the gravity side\cite{progress}. \\\\
 Our results in this work and \cite{Dadras22} are based on the general fact that by coupling both sides of TFD its entanglement entropy will change. This would imply that the ``duality'' between a black hole and a lower dimensional geometry exists in higher dimensions too, although it may be more complicated (See for example \cite{Duffin18} for finite temperature CFT in higher dimensions). \\\\
\textbf{Acknowledgment} ~ I am grateful to John Preskill for helpful discussions and especially to Alexei Kitaev for helpful discussions and advice at different stages of this project. I am also grateful to M.M. Sheikh-Jabbari for helpful discussions on the second version of the paper.  This work is supported by Simons Foundation, as well as by the Institute of Quantum Information and Matter, the NSF Frontier center funded in part by the Gordon and Betty Moore Foundation.
\appendix
\section{The Gibbons-Hawking-York action close to the boundary of AdS$_3$  \label{appendixA}}
	In this section, we will study the dynamics of the ``physical boundary'' in the vacuum asymptotically AdS$_3$ $(\AAdS_3)$ spacetime \cite{Ashtekar00,Ashtekar84}. To define the physical boundary, we first briefly review two important properties of the $\AAdS_3$ spacetime $(\calM, g_{\mu\nu})$: \\\\
	$(i)$ ~ The spacetime $(\calM, g_{\mu\nu})$ satisfies Einstein's equation.
		\be
R_{\mu\nu} - \frac{1}{2} R g_{\mu\nu} + \Lambda g_{\mu\nu} = 0.
\ee
$(ii)$ There is a spacetime $(\wtilde \calM, \wtilde g_{\mu\nu})$ with boundary $\wtilde \calI$ and a smooth function $\Omega$ such that  $\wtilde g_{\mu\nu} = \Omega^2 g_{\mu\nu}$ and $\wtilde \calM - \wtilde \calI$ is diffeomorphic to $\calM$. Furthermore, $\Omega \Big|_{\wtilde \calI} = 0$ with $\nabla_\mu \Omega \Big|_{\wtilde \calI} \neq 0$, and the topology of $\widetilde I$ is $S \times R$. \\\\
The second condition allows us to choose the function $\Omega$ locally as one of the coordinates (denoted by $r$). In this coordinate system, components of the metric will be
\be
 g_{rr},~~~~ g_{r i},~~~~~ g_{ij},~~~~~~ i,j = 1,2,
 \ee
  and it locally defines a decomposition of the spacetime by $r= \const$ co-dimension one hypersurfaces. We define the physical boundary as the timelike hypersurface $r = \vepsilon \ll 1$. We will choose the coordinate system $r,t,\phi$ so that $dr \wedge dt \wedge d\phi$ is a positive volume form. $S_{GHY}$ takes the following form,
  \be \label{HG}
  \begin{aligned} &
  S =  -\frac{1}{8\pi G} \int_{r=\vepsilon}\sqrt{-\det g_{ij}} ~  \Big(K+1\Big) ~ du \wedge dv.
  \end{aligned}
  \ee
 We are going to study the dynamics associated with the above action. The action becomes divergent as we approach the boundary. However, we will give a natural way to regularize the action.  \\\\ 
   To compute the GHY term, condition (ii) clearly implies that we only need to compute $K$ to second order in $r$.
  From the first condition, the dynamics of the boundary (in the absence of the matter) should satisfy Hamiltonian and momentum constraints:
\be \label{Hc}
\begin{aligned} &
^2R + K^{ij} K_{ij} -K^2 - 2\Lambda = 0, \\&
D^b T_{ab} = 0,~~~~~~T_{ab} = K_{ab} - K h_{ab}- h_{ab} \end{aligned} 
\ee
with $D$ to be the induced covariant derivative on the physical boundary. In particular, we assume $T_{ab}$ remains finite as we send $\vepsilon \rightarrow 0$ .
 On the physical boundary, these two equations can be studied perturbatively as a function of $r$. However, it is practically easier to study them in $\wtilde \calM$ close to $r = 0$ where the boundary $\wtilde {\calI}$ is located. Note that the one form associated with the normal vector to the hyper-surfaces $r = \const$ is $n_{\mu} = (g^{rr})^{-1/2} (1,0,0)$. We have the following relations, 
 \be 
 K_{ij} = r^{-1} \wtilde K_{ij} - \frac{\wtilde n^r}{r^2} ~ \wtilde g_{ij},~~~~~~^2R = r^2 ~(^2\wtilde R).
 \ee
  Here, $n$ is the inward normal vector, and so $K_{ij} = \langle \nabla_i n, \p_j \rangle$.
The constraint \ref{Hc} can be rewritten as 
\be \label{cHc}
r^2 \Big( \wtilde K_{ij} \wtilde K^{ij} - \wtilde K^2 \Big) - 2 \Big( \wtilde n^r \Big)^2 + 2r \wtilde K \Big( \wtilde n^{\,r}  \Big) + r^2 ~^2\wtilde R + 2 = 0.
\ee
 The leading order implies that $\wtilde n_{(0)}^r = 1$, which means the boundary is timelike.
Close to the boundary $\calI$, $\wtilde n^r, \, \wtilde g_{ij} \, \wtilde K_{ij}$ can be expanded as
\be
\wtilde K_{ij} = \wtilde K^{(0)}_{ij} + r \, \wtilde K^{(1)}_{ij} + O(r^2),~~~~\wtilde g_{ij} = \wtilde g^{(0)}_{ij} + r \, \wtilde g^{(1)}_{ij} + O(r^2),~~~~ \wtilde n^r = 1 + r ~ \wtilde n_{(1)}^{r} +r^2 ~ \wtilde n_{(2)}^{r}+O(r^3).\ee Plugging back into \ref{cHc}, to the second order in $r$ we have:
\be
\wtilde K^{(0)}_{ij} \wtilde K^{ij(0)} - \frac{1}{2} \wtilde K_0^2 - 4 \wtilde n^r_{(2)} + 2 (\wtilde K^{(0)}_{ij} \wtilde g^{ij(1)} +\wtilde K^{(1)}_{ij} \wtilde g^{ij(0)}) +~ ^2\wtilde R^{(0)} =0,~~~~~ \wtilde n_{(1)}^{\,r} = \frac{\wtilde K^0}{2}\ee
Moreover, $\wtilde K^{(0)}_{ij} = \frac{1}{2} \wtilde K^{(0)} \wtilde g^{(0)}_{ij}$. This can be seen, for example, from the finiteness of the Brown-York tensor \cite{Brown92,Bala99} in the limit $r \rightarrow 0$
\be
\begin{aligned} &
K_{ab} - K h_{ab} -h_{ab} = \frac{\tilde K_{ab}}{r} - \frac{\tilde n^r}{r^2} \tilde h_{ab} -\Big( r \tilde K - 2\tilde n^r \Big) \frac{\tilde h_{ab}}{r^2} - \frac{\tilde h_{ab}}{r^2} \\& = \frac{\tilde K_{ab} - \tilde K \tilde h_{ab}}{r} +\frac{\tilde n^r - 1}{r^2} \tilde h_{ab} = \frac{\tilde K^0_{ab} - \tilde K^0 \tilde h^0_{ab}}{r} -\frac{\tilde n^1}{r} \tilde h^0_{ab} +O(1) = \frac{\tilde K^0_{ab} - \frac{\tilde K^0}{2} \tilde h^0_{ab}}{r} +O(1),\end{aligned} 
\ee
which yields:
\be \label{Bcurve}
\frac{1}{2}~ ^2\wtilde R^{(0)} = 2 \wtilde n^r_{(2)} - (\wtilde K^{(0)}_{ij} \wtilde g^{ij(1)} +\wtilde K^{(1)}_{ij} \wtilde g^{ij(0)}). \ee
On the other hand, we can rewrite the trace of the extrinsic curvature as follows:
 \be \label{Extr}
K = -2 -r \Big( 2 \, \wtilde n^{r\,(1)}- \wtilde g^{ij(0)} \wtilde K^{(0)}_{ij} \Big) - r^2 \Big( 2\wtilde n^{\,r(2)} -\wtilde K^{(0)}_{ij} \wtilde g^{ij(1)}- \wtilde K^{(1)}_{ij} \wtilde g^{ij(0)} \Big).\ee 
Comparing \ref{Bcurve} and \ref{Extr} renders
\be \label{extrinsic}
K = -2 - \frac{1}{2} r^2 ~^2\wtilde R^{(0)} + O(r^3).
\ee
However, the $\wtilde R^{(0)}$ term above does not lead to any dynamics. The reason is that if we plug the above in $S_{GHY}$, the contribution of this term will be through
\be
\frac{1}{16\pi G} \int ~ d^2 x ~ \sqrt{-\det\wtilde g_{ij}^{(0)}} ~~ ^2{\wtilde R^{(0)}}, 
\ee
which is topological and may contribute to the zero temperature entropy. Therefore, a possible dynamics only comes from the determinant of the boundary metric.

\bibliography{BTZ}
\bibliographystyle{JHEP}

\end{document}